\documentclass[fleqn,twoside]{article}
\usepackage{amsmath}
\usepackage{espcrc2}
\usepackage{psfrag}
\usepackage{graphicx}

\newcommand{\op}[1]{\mathbf{#1}}
\newcommand{\av}[1]{\left\langle#1\right\rangle}

\title{Connected Correlators in Random Geometries}

\author{
J.~Ambj\o rn %
\address{The Niels Bohr Institute, Blegdamsvej 17, 
         DK-2100 Copenhagen \O, Denmark\\
$^{\rm b}\;$Fakult\"{a}t f\"{u}r Physik, Universit\"{a}t 
            Bielefeld, 335615 Bielefeld, Germany \\
$^{\rm c}\;$Institute of Comp.\ Science, 
            Jagellonian  University, 30-072 Krakow, Poland\\
$^{\rm d}\;$Institute of Physics, 
            Jagellonian University, 30-059 Krakow, Poland},
P.~Bialas$^{\rm b,c}$, J.~Jurkiewicz$^{\rm a,d}$} 

\begin{document}

\begin{abstract}
We analyze correlation functions in a toy model of a random geometry
interacting with matter.  We show that in general the connected
correlator will contain a long--range scaling part which is in some
sense a remnant of the disconnected part.  This result supports the
previously conjectured general form of correlation functions. We
discuss the interplay between matter and geometry and the role of
the symmetry in the matter sector.
\end{abstract}

\maketitle

\section{Introduction}

Simplicial quantum gravity, also called dynamical triangulations (DT),
provides a simple and constructive definition of the path integral
over the geometries which enables the use of Monte--Carlo techniques
(see \cite{Thorleifsson:1998jr} and references therein) and
gives an opportunity to make a fully non--pertubative study
of various quantities including correlation functions.

A possible reparametrization invariant definition of a two--point
correlator in the canonical ensemble is \cite{deBakker:1995he}~:
\begin{equation}\label{dis}     
G^{\op{AB}}_n(r)=\Big\langle\frac{1}{n}\sum_{i,j}^n \op{A}_i\op{B}_j
\delta_{d(i,j),r}\Big\rangle_n
\end{equation}
The average is taken over all configurations with a fixed volume ($n$).
Symbol $d(i,j)$ denotes the geodesic distance -- the shortest path
between the points $i$ and $j$.

The definition of the connected correlation functions is more
ambiguous \cite{deBakker:1995he,Bialas:1997ei}.  Here we use the
following formula \cite{Ambjorn:1998vd}~:
\begin{multline} \label{con}   
G_n^{\op{AB}conn}(r)=\\
\kern-2mm\Big\langle\frac{1}{n}\sum_{i,j}^n \delta_{d(i,j),r}(\op{A}_i-\langle \op{A}\rangle_n)(\op{B}_j-\langle \op{B}\rangle_n)\Big\rangle_n.
\end{multline}

The correlators (\ref{dis}) are very difficult to study and
very few analytic results exist \cite{Ambjorn:1998vd,Kawai:1993cj}.  A
growing body of numerical evidence suggests however the existence of a
simple structure common to the correlators (\ref{dis}) in various
ensembles of random geometries \cite{Ambjorn:1998vd}.

In this contribution we analyze a simple model of random geometry
interacting with non--critical matter which permits the detailed
analytic analysis of the correlators (\ref{dis}) and (\ref{con}). The
importance of such a study stems from the fact that finite size
effects can make the non--critical correlations appear as long--range and
this effect must be properly subtracted \cite{Ambjorn:1998vd}.

\section{Branched Polymers}

Branched Polymers (BP) provide a very simple but non--trivial example
of a random geometry ensemble \cite{Ambjorn:1986dn}. Moreover they
exhibit a wide range of properties in common with  higher
dimensional DT systems \cite{Bialas:1996eh}. 

Here we consider the Ising model on a BP as the simplest model of matter
coupled to a random geometry.  The system is described by two coupled
equations~\cite{Ambjorn:1993rp}:
\begin{equation}\begin{split}\label{sys}
Z_{\pm 1}(\mu) &= 
e^{-\mu}\sum_{s=\pm 1}
e^{ s h} e^{ \pm s\beta }F\bigl(Z_{s}(\mu)\bigr)
                \end{split}
\end{equation}
where $\beta$ is the inverse temperature and $h$ the magnetic field.
For simplicity we chose $F(z)=1+z+z^2$ which corresponds to a BP with
$q_i\le 3$ where $q_i$ denotes a number of branches in the vertex $i$.  For
this choice the system is always in the generic (elongated) phase with
$\gamma_{str}=\frac{1}{2}$ and $d_H=2$.

The general form of the correlation function (\ref{dis}) in this model 
is \cite{gen}~:
\begin{multline}\label{Isingdis}        
G^{\op{AB}}_n(r)\approx 
2 \sqrt{ n}\, \mathcal{C}
\av{\op{A}}\av{\op{B}}\,
g\bigl(a_0 \frac{r+\delta^{(0)}_\op{A}+\delta^{(0)}_\op{B}}{2\sqrt{n}}\bigr)\\
+ 2\sqrt{ n}\,
A_1 B_1\,e^{-\mu_1 r}\, g\bigl(a_1 \frac{r+\delta^{(1)}_\op{A}+\delta^{(1)}_\op{B}}{2\sqrt n}\bigr)  
\end{multline}
where $g(x)=x e^{-x^2}$ and the non--negative parameters 
$\mathcal{C}$, $a_0$, $a_1$, $\mu_{1}$ depend only on $\beta$ and $h$
and are independent of the  choice of operators $\op{A}(\op{B})$.  

From (\ref{Isingdis}) and (\ref{con}) it follows that~:
\begin{multline}\label{Isingcon}        
G^{\op{AB}conn}_n(r)\approx 
\frac{1}{2 \sqrt{ n}}\, 
X_{\op{AB}}\,
g''\bigl(a_0 \frac{r+\delta^{(0)}_\op{AB}}{2\sqrt{n}}\bigr)\\ + 
2\sqrt{ n}\,
Y_{\op{AB}}\,e^{-\mu_1 r}\, g\bigl(a_1 \frac{r+\delta^{(1)}_\op{AB}}{2\sqrt n}\bigr)
\end{multline}
The parameters $X_{\op{AB}}$, $Y_{\op{AB}}$ and $\delta^{(k)}_{\op{AB}}$ 
appearing
in (\ref{Isingcon}) are simple functions of the parameters in the formula
(\ref{Isingdis}).

When $h=0$  symmetry implies $Z_{+}(\mu)=Z_{-}(\mu)$ and
the equations (\ref{sys}) decouple \cite{Ambjorn:1993rp}. The model is
solvable and we obtain~:
\begin{equation}\label{sym}     
X_{\op{ss}}=0\quad \text{and}\quad Y_{\op{qq}}=0                
\end{equation}
We see that in this special case the matter--matter correlator (\ref{con}) 
contains only the short--range part, while the purely geometric correlator
appears as a long--range one.

In case of the non--zero magnetic field the system (\ref{sys}) cannot be
solved analytically and the values of the coefficients $X_{\op{AB}}$
and $Y_{\op{AB}}$ in  (\ref{Isingcon}) must be evaluated numerically.  
We performed the calculations for  $\beta=1.0$ and $h=0.1$ corresponding to
$\mu_{1}\approx0.5605$.  We compared the analytical predictions with the MC
simulations of systems with $n$ equal to 250, 1000 and 4000 to check
the validity of the large $n$ approximation.

\begin{figure}[t]
\begin{center}
\psfrag{r}[t][b][1][0]{$r$}
\psfrag{x}[t][b][1][0]{$x$}
\includegraphics[width=7.2cm,bb=76 140 515 434]{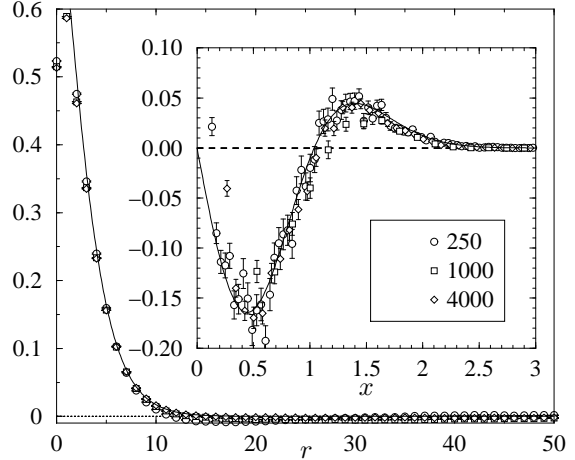}        
\end{center}
\caption{\label{ssc}Spin--spin correlation function.}
\end{figure}

\begin{figure}[t]
\begin{center}
\psfrag{r}[B][B][1][0]{$r$}
\psfrag{x}{$x$}
\includegraphics[width=7.2cm,bb=66 140 511 434]{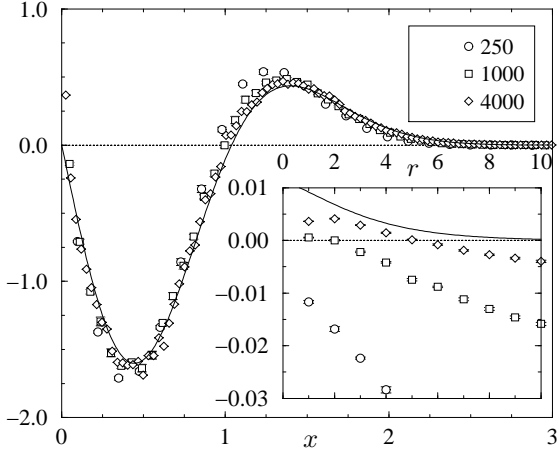}        
\end{center}
\caption{\label{qqc}``Curvature--curvature'' ($q$--$q$) correlation function.}
\end{figure}

In the figures \ref{ssc} and
\ref{qqc} we  present the results for the connected spin--spin and
curvature--curvature correlations. 
As predicted each contains a long--range scaling and a
short range non--scaling part. They retain however some ``memory'' of
the symmetric case (\ref{sym})~: the matter (spin--spin) correlation
function is  dominated by the non--scaling part and the geometric
correlator is dominated by the scaling--part so that the  structure
(\ref{Isingcon}) is not immediately visible.

To make the structure of (\ref{Isingcon}) more explicit we redraw
in the inlay in figure~\ref{ssc}
the spin--spin correlators for various sizes rescaled by $2\sqrt{n}$ as a
function of the scaling variable $x=\frac{r}{2\sqrt{n}}$. 

In figure~\ref{qqc} we again emphasize the scaling behavior
by plotting the rescaled geometric correlators on top of each other. 
The non scaling part can only be seen for small $r$ and large $n$ 
and  is shown in the inlay. 
 
In both cases the MC data are consistent with the  analytic predictions
shown by the continuous line.

\section{Improved correlator}

The form (\ref{Isingdis}) of the correlation function suggests an improved
definition of the connected correlator. 
For  $r\gg \frac{1}{\mu_1}$ we have
\begin{equation}\label{GABassympt}      
G^{\op{AB}}(r)\approx \av{\op{A}}\av{\op{B}}
G^{\op{11}}\bigl(r+\delta^{eff}_{\op{A}} +\delta^{eff}_{\op{B}} \bigr)
\end{equation}
with $\delta^{eff}_{\op{A}}=\delta^{(0)}_\op{A}-\delta^{(0)}_\op{1}$. 
This relation was the source of the scaling term in the connected correlators
(\ref{con}).
It is possible to subtract this term by redefining the connected
correlator to~\cite{Ambjorn:1998vd}:
\begin{multline}\label{GABimp}  
G^{\op{AB}conn}_n(r)=
G^{\op{AB}}_n(r)\\
-\av{\op{A}}_nG^{\op{1B}}(r+\delta^{eff}_\op{A})
-\av{\op{B}}G^{\op{1A}}_n(r+\delta^{eff}_\op{B})\\ 
+\av{\op{A}}\av{\op{B}}
G^{\op{11}}_n(r+\delta^{eff}_\op{A}+\delta^{eff}_\op{B})
\end{multline}
The values of shifts $\delta^{eff}$ can be obtained 
by fitting the relation (\ref{GABassympt}) to the functions 
$G^{\op{1A}}$ and $G^{\op{1B}}$ \cite{Ambjorn:1998vd}. 

After inserting  (\ref{dis}) into (\ref{GABimp}) the scaling
term in (\ref{con}) will decrease by a factor $\frac{1}{n}$
while the  non--scaling term will retain its leading order behavior. 

\section{Discussion}

The obtained results
can be generalized to a BP model with other types of matter and with
two--point geometric interactions \cite{gen}. The main
result remains unchanged~: the correlator (\ref{Isingcon}) contains
scaling and non--scaling parts,
but  additional non--scaling terms appear. The result (\ref{sym}) can
also be generalized. If the matter field takes values in a group and
the partition function is symmetric under the action of this group
then $X_{AB}=0$ for all matter operators (respecting the symmetry).

Numerical simulations indicate that the asymptotic behavior
(\ref{GABassympt}) is even more general and valid in almost every kind
of random geometry ensembles.  In particular the behavior of the Ising
model in a magnetic field coupled to the 2D simplicial quantum gravity exhibits
qualitatively the same features and the improved correlator can be
successfully used \cite{Ambjorn:1998vd}. Similar structure is observed
for the correlators of the local action density
of the  Abelian gauge fields in the
4D simplicial quantum gravity \cite{Ambjorn:1999ix}.

It seems that there exists a high degree of ``universality'' in the
structure of correlators in systems with a random geometry. Simple
models as the one described above can give us so far the unique opportunity
to study this structure analytically. 
 
The picture which emerges is the following~: the purely geometric
correlators (\ref{con}) behave asymptotically according to
(\ref{GABassympt}). This leads to the appearance of the long--range
correlation term in the disconnected correlator (\ref{dis}).  The
matter fields pick up these correlations through the {\em local}
coupling to the geometry.  Where  symmetry is present this local
coupling is inhibited and the long--range term is absent.
 
{\bf Acknowledgments}
P.~B. was supported by the Alexander von Humboldt foundation and TMR network
ERBFMRX-CT97-0122.    

\bibliographystyle{utphys2} 
\bibliography{/home/pbialas/BIB/grav}

\end{document}